\documentclass{ws-procs9x6}

\begin{document}

\title{Correlations between $<\lowercase{p}_T>$ and multiplicity in a single BFKL Pomeron
\footnote{{\sf \uppercase{T}alk given by \uppercase{C}. \uppercase{M}erino.}}
\footnote{{\sf \uppercase{C}ontribution to the \uppercase{P}roceedings of the \uppercase{XXXII} 
\uppercase{I}nternational \uppercase{S}ymposium
on \uppercase{M}ultiparticle \uppercase{D}ynamics (\uppercase{ISMD}), \uppercase{A}lushta, 
\uppercase{C}rimea (\uppercase{U}kraine), 7-13 \uppercase{S}eptember 2002, \uppercase{W}orld
\uppercase{S}cientific.}}}

\author{M.A. BRAUN}

\address{ Dept. High-Energy Physics\\
St. Petersburg State University\\ 
198504 St. Petersburg, Russia\\
E-mail: Braun1@pobox.spbu.ru}  

\author{C. MERINO and G. RODRIGUEZ}

\address{Department of Particle Physics, Facultade de F\'\i sica, \\
Universidade de Santiago de Compostela, \\
Campus Universitario s/n, \\
Santiago de Compostela, Galice, Spain\\
E-mail: Merino@fpaxp1.usc.es\\
E-mail: Grod@fpaxp1.usc.es}


\maketitle

\abstracts{Strong correlations are obtained between the number and the average 
transverse momentum of jets 
emitted by the exchange of a single BFKL Pomeron.}

\section{Introduction}

Strong correlations are observed experimentally
between the average
$p_T$ and multiplicities of particles produced in high-energy hadronic
collisions [1]. Average $p_T$ grows with multiplicity. To interpret this fact
it is tacitly assumed that with only
one hard collision
there are no correlations between $<p_T>$ and multiplicity. Theoretically
this assumption can only be tested within the 
Balitskii-Fadin-Kuraev-Lipatov (BFKL) dynamics, which presents a
detailed description of particle (actually jet) production at high energies
under certain simplifying assumptions (a fixed small coupling constant).
The present calculation is aimed to see if there exist correlations between
$<p_T>$ and the number of produced jets in the hard Pomeron described by the BFKL
chain of interacting reggeized gluons~[2]. We limit ourselves to the leading order
BFKL model.

\section{The Formalism}

The BFKL equation for the amputated BFKL amplitude, $f(y,k)$,
when $y$ is the rapidity and $k$ is the two-dimensional transverse
momentum of the virtual (Reggeized) gluon, may be written in the form
\begin{equation}
f(y,k)=f^{(0)}(y,k)+\bar{\alpha}_s\int_0^ydy_1\int\frac{d^2k_1}{\pi q^2}
\Big(\frac{k^2}{k_1^2}f(y_1,k_1)-f(y,k)\theta (k^2-q^2)\Big),
\label{eq:paper2}
\end{equation}
where 
$\bar{\alpha}_s=3\alpha_s/\pi$ and $q=k-k_1$ is the transverse
momentum of the emitted (real) gluon.

Defining as an observable jet a real gluon with $q^2\geq~\mu^2$, one
splits the integration over momenta and thus the integration kernel in~(\ref{eq:paper2}) 
into two parts, a resolved one, $K_R$,
corresponding to emitted gluons with $q^2>~\mu^2$, and an unresolved one, $K_{UV}$,
which combines emission of gluons with $q^2<~\mu^2$ and the subtraction term 
in~(\ref{eq:paper2}). Exclusive probabilities to produce $n$ jets are obtained by
introducing $n$ operators $K_R$ between the Green functions of the BFKL
equations with kernel $K_{UV}$~[3]. If one presents the full gluon distribution
$f$ as a sum of contributions $f_n$ from the production of $n$ jets
then one gets  a recursive relation
\begin{equation}
f_n(y)=\int_0^ydy_1K(y-y_1)f_{n-1}(y_1),
\label{eq:paper8}
\end{equation}
where $K(y)$ is an $y$-dependent operator in the transverse momentum space
\begin{equation}
K(y)=e^{yK_{UV}}K_R.
\label{eq:paper9}
\end{equation}
Eq.~(\ref{eq:paper8}) allows one to successively calculate the relative probabilities to
produce $n=0,1,2,...$ jets starting from the no-jet contribution.

The exclusive physical probabilities to observe $n$ jets are obtained by convoluting $f_n$ 
with the gluon distribution
in the projectile (the projectile impact factor). 
Both the impact factors of the target and
of the projectile should vanish as
$k\rightarrow 0$.

\section{The Calculation}

We are interested in the average values of $<q>_n$ in the observed jets,
provided their number $n$ is fixed. The momentum
$k$ which serves as an argument of $f(y,k)$ refers to the virtual gluon,
and not to the emitted one, whose momentum $q$ is hidden inside the kernel
$K_R$. Therefore to find an average of any quantity $\phi(q)$ depending
on the emitted real jet momentum, one has to introduce the function $\phi(q)$
into the integral defining $K_R$, thus changing the kernel $K_R$ to the kernel $K_{av}$:
\begin{equation}
\Big(K_{av}f\Big)(k)=\bar{\alpha_s}k^2\int\frac{d^2k_1}{\pi q^2 k_1^2}
\theta(q^2-\mu^2)\phi(q)f(k_1).
\label{eq:paper14}
\end{equation}
With $n$ jets, one has to substitute
one of the $n$ operators $K_R$ which generate the jets by $K_{av}$, take a sum of all
such substitutions, and divide by $n$. One has further to integrate over
all momenta of the virtual gluon $k$ multiplied by the projectile impact factor,
and normalize the result to the total probability to have $n$ jets. To formalize
this recipe we introduce a generalized operator in the virtual gluon momentum space
\begin{equation}
K_1(y)=e^{yK_{UV}}[K_R+K_{av}].
\label{eq:paper15}
\end{equation}
Let the function $F(y,k)$ obey the equation
\begin{equation}
F(y)=f_0(y)+\int_0^ydy_1K_1(y-y_1)F(y_1).
\label{eq:paper16}
\end{equation}
One can split the function $F$ into a sum of contributions $F_{nm}$
corresponding to the action of  $n$ operators $K_1$, out of which
$m=0,1,...n$, are operators $K_{av}$
(evidently $F_{n0}=f_n$). We are interested in the contribution
$F_{n1}\equiv g_n$ which contains a single operator $K_{av}$. The average
value of interest
is determined by
\begin{equation}
<\phi(q)>_n=\frac{1}{n}\frac{\int (dk^2/k^4)h(k)g_n(y,k)}
{\int (dk^2/k^4)h(k)f_n(y,k)}.
\label{eq:paper18}
\end{equation}

In analogy with Eq.~(\ref{eq:paper8}), one easily sets up a recursion relation for $g_n$:
\begin{equation}
g_n=\int_0^ydy_1K(y-y_1)g_{n-1}(y_1)+\int_0^ydy_1
e^{(y-y_1)K_{UV}}K_{av}f_{n-1}(y_1),
\label{eq:paper19}
\end{equation}
with the initial condition $g_0(y)=0$. Together with~(\ref{eq:paper8}), this relation
allows one to calculate the function $g_n$ for $n=1,2,...$, and then to use~(\ref{eq:paper18}) to
find the desired averages.

The concrete choice of $\phi(q)$ is restricted by the condition of
convergence at large $q$: $\phi(q)<q^2$, as $q\rightarrow\infty$.
To facilitate our calculation we make a natural choice
$\phi(q)=q$, which allows the angular integration to be done analytically.   

\section{The Results}

We defined our jets by taking $\mu=2$ GeV/c. As for the cutoffs,
we used
\begin{equation}
1 GeV/c < k_1 < 100 GeV/c,
\label{eq:paper3}
\end{equation}
and we used a simplified expression for the virtual photon impact factor,
independent of rapidity~[2].

We have calculated the functions $f_n$ and $g_n$ from Eqs.~(\ref{eq:paper8}) 
and (\ref{eq:paper19})
up to $n=5$ and $y=15$. Following [3]
we have used the expansion in $N$ Chebyshev polynomials to discretize the
kernels in a simple way.

\begin{figure}[ht]
\centerline{\epsfxsize=3.93in\epsfbox{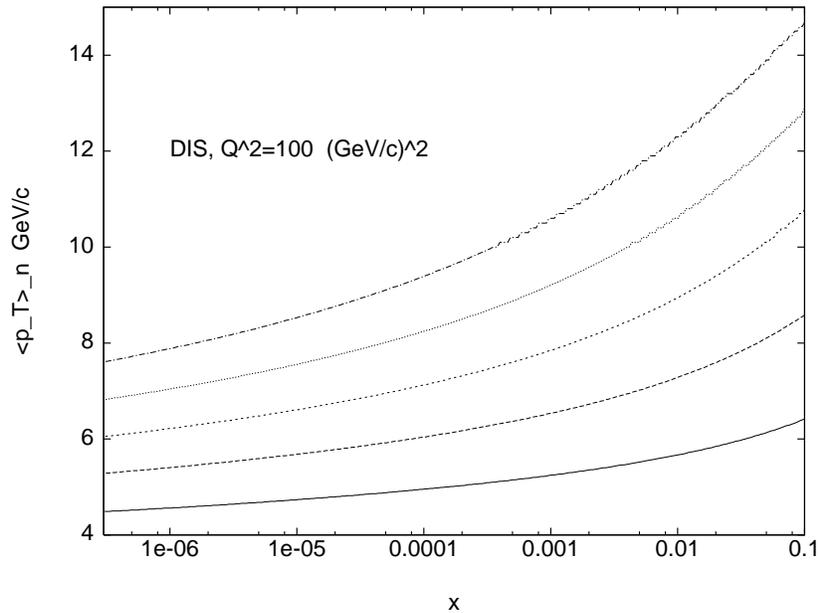}}   
\caption{Average $<p_T>_n$  for a  fixed number
$n$ of jets produced in $\gamma*$-hadron collisions, as a function of
$x$  at $Q^2=100$ (GeV/c)$^2$. Curves from bottom to top
correspond to $n=1,2,...5$.
\label{figure2}}
\end{figure}

In Figure~\ref{figure2} we present the averages $<q>_n$
for $n=1-5$ and $x=e^{-y}=~3.10^{-7}-0.1$,
for the $\gamma^*$-hadron collisions (DIS) at $Q^2~=~100$~(GeV/c)$^2$.

As one observes, $<q>_n$ strongly grows with $n$ at all
rapidities, being the growth approximately linear.

As an interesting by-product of our study we find that the averages $<q>_n$
go down with rapidity for all $n\geq 2$. This is quite unexpected, since
in the BFKL approach an overall average $<q>$ rapidly
grows with $y$.

Similar results are obtained for purely hadronic collisions~[2].

\section{Discussion} 

Emissions of
high-$p_T$ jets in DIS seem to be a suitable
place to
see the BFKL signatures. Our results show that in such emissions strong
positive correlations are predicted between $<p_T>$ and the number of jets,
already for a single Pomeron exchange. This indicates that in fact such
correlations are
already present in the basic mechanism of jet production. 
The linear growth of $<p_T>$ with $n$ that has been obtained could be a
random-walk effect, $<p_T>$ becoming larger
and larger at each step (with each new produced jet) [4]. 
The extension of our study to the case of the BFKL
equation with a running coupling constant would be important in order to
stablish the stability of our results.

An unexpected result obtained in our calculation is
that $<q>_n$ at fixed $n\geq 2$ fall with energy.
Certainly this phenomenon deserves further investigation
including higher $y$ and/or $n$. We hope that it can be tested experimentally
as a possible signature of the BFKL Pomeron.

\section*{Acknowledgments}

This work is supported by CICYT (Spain), FPA2002-01161,
and by the RFFI grant 01-0-17137 (Russia).

\section*{References}

\noindent
1. UA1 Collaboration, C. Ciapetti in {\it The Quark Structure of
Matter}, edited by
M. Jacob and K. Winter (1986), p. 455;
F. Ceradini, Proceedings of the International Europhys. Conference on High-Energy Physics,
Bari, edited by L. Nitti and G. Preparata (1985), and references therein.

\noindent
2. M.A. Braun, C. Merino, and G. Rodr\'\i guez, {\it Phys. Rev.} {\bf D 65},
114001 (2002).

\noindent
3. J. Kwiecinski, C.A.M. Lewis, and A.D. Martin, {\it Phys. Rev.} {\bf D 54},
6664 (1996).

\noindent
4. We thank A.B. Kaidalov for enlightening discussions on this point.



\end{document}